\documentclass[aps,nofootinbib,showpacs,preprintnumbers,amsmath,amssymb]{revtex4}
\usepackage[utf8]{inputenc}
\def\be{\begin{equation}}
\def\ee{\end{equation}}
\def\bea{\begin{eqnarray}}
\def\eea{\end{eqnarray}}
\def\bear{\begin{array}}
\def\eear{\end{array}}
\newcommand{\MSbar}{\overline{\rm MS}}  

\newcommand{\A}{{\mathcal{A}}}

\usepackage{graphics}
\usepackage{graphicx}
\usepackage{dcolumn}
\usepackage{bm}
\usepackage{epsfig}
\usepackage{graphicx}
\usepackage{multirow}
\usepackage{dcolumn}
\usepackage{graphicx,epsfig}%

\begin{document}

\preprint{USM-TH-319}
\title{Gluon Propagator in Fractional Analytic Perturbation Theory} 
\author{Pedro Allendes}
\author{C\'esar Ayala}
 \email{c.ayala86@gmail.com}
\author{Gorazd Cveti\v{c}}

\affiliation{Department of Physics, Universidad T{\'e}cnica Federico
Santa Mar{\'\i}a (UTFSM), Valpara{\'\i}so, Chile\\
}

\date{\today}

\begin{abstract}
We consider the gluon propagator in the Landau gauge at low
spacelike momenta and with the dressing function $Z(Q^2)$
at the two-loop order. 
We incorporate the nonperturbative effects by making the (noninteger)
powers of the QCD coupling in the dressing function $Z(Q^2)$
analytic (holomorphic) via the
Fractional Analytic Perturbation Theory (FAPT) model, and simultaneously
introducing the gluon dynamical mass in the propagator as motivated 
by the previous analyses of the Dyson-Schwinger equations.
The obtained propagator has behavior compatible with the unquenched 
lattice data ($N_f=2+1$) at low spacelike momenta 
$0.4 \ {\rm GeV} < Q \alt 10$ GeV. 
We conclude that the removal of the unphysical Landau singularities of the 
powers of the coupling via the (F)APT prescription, in conjunction with the
introduction of the dynamical mass $M \approx 0.62$ GeV 
of the gluon, leads to an
acceptable behavior of the propagator in the infrared regime.

\end{abstract}

\maketitle

\section{Introduction}
\label{sec:intr}

The perturbative approach in QCD (pQCD) is known to work well at high
momenta ($|q^2| \agt 10 \ {\rm GeV}^2$). However, at low momenta  ($|q^2| \sim 1 \ {\rm GeV}^2$)
it is unreliable, principally because the pQCD coupling parameter
$a_{\rm pt}(Q^2) \equiv \alpha_s(Q^2)/\pi$  has (Landau) singularities
at spacelike low momenta $q$: $0 < Q^2 (\equiv - q^2)  \alt 10^{-1} \ {\rm GeV}^2$.
This singularity structure is incompatible with the analyticity
properties of the (dimensionless) spacelike physical quantities ${\cal D}(Q^2)$ such as
(derivatives of) current correlators, structure functions, propagator
dressing functions, etc.
By the general principles of the 
quantum field theory (QFT) \cite{BS,Oehme}, such physical quantities must be 
analytic (holomorphic) functions in the entire
complex $Q^2$-plane except on the negative semiaxis, i.e.,
$Q^2 \in \mathbb{C} \backslash (-\infty, -M^2_{\rm thr}]$, where
$M_{\rm thr} \sim 10^{-1}$ GeV is a threshold mass typical of the hadronic sector.
Therefore, if these quantities ${\cal D}(Q^2)$ are to be evaluated  
as functions of $a(\kappa Q^2)$  (where $\kappa \sim 1$ is the renormalization scale
parameter), the coupling $a(\kappa Q^2)$ should have qualitatively the same
analiticity properties as ${\cal D}(Q^2)$.

The first model with holomorphic QCD coupling was constructed in 
\cite{ShS,MSS,Sh}. In this model, named also Analytic Perturbation Theory (APT),
the coupling $\A(Q^2)$ [the analytic analog of $a(Q^2)$] is obtained
from $a(Q^2)$ by a minimal analytization approach. Namely,
$\A(Q^2)$ is written as a dispersion integral involving the
discontinuity function $\rho(\sigma) \equiv  {\rm Im} \A(-\sigma - i \epsilon)$,
where $\rho(\sigma)$ is taken equal to its
perturbative counterpart $\rho^{\rm (pt)}(\sigma) \equiv {\rm Im} a(-\sigma - i \epsilon)$
on the entire negative axis in the $Q^2$-plane ($\sigma > 0$), while the
Landau singularities of $a(Q^2)$ on the positive $Q^2$-axis
were removed: $\rho(\sigma)=0$ for $\sigma < 0$. Later on, the APT model was extended to
analytic analogs $\A_{\nu}(Q^2)$ of the powers $a(Q^2)^{\nu}$
(for $\nu$ real, in general noninteger) in the works \cite{BMS1,BMS2,BMS3,BP}
(for reviews, see \cite{Bakulev}). In these works, at the one-loop level 
of the underlying pQCD an explicit expression of $\A_{\nu}^{(1-\ell.)}(Q^2)$ 
was obtained; at higher loop levels, the coupling $\A_{\nu}$ was expressed in terms of
a series in $\nu$-derivatives of $\A_{\nu}^{(1-\ell.)}(Q^2)$. This extended model
is called Fractional Analitic Perturbation Theory (FAPT). 

Other models of analytic QCD were constructed later on.
One such model is obtained by using the method of minimal analytization
for the function $d \ln a(Q^2)/d \ln Q^2$, Refs.~\cite{Nesterenko1}, 
which leads to a holomorphic coupling which diverges at $Q^2=0$.
Another such model is, for example, the two-delta analytic
QCD model \cite{2danQCD} which uses for the discontinuity function 
$\rho(\sigma) \equiv {\rm Im} \A(-\sigma - i \epsilon)$
the underlying pQCD values at high $\sigma$ 
and parametrizes $\rho(\sigma)$ at low $\sigma$ 
with two delta functions, and at high $Q^2$ the coupling
$\A(Q^2)$ practically coincides with the pQCD coupling $a(Q^2)$.
Analytization of general powers $a(Q^2)^{\nu}$ in general analytic
QCD models of $\A(Q^2)$ was performed in Ref.~\cite{GCAK}.

In this work we are using FAPT to evaluate the dressing function $Z(Q^2)$
of the gluon propagator
at low spacelike momenta in the Landau gauge, at the two-loop level of the
underlying ($\MSbar$) pQCD. An analysis of this quantity
was performed in Ref.~\cite{Magradze:1998ng} in the context of APT; 
however, at that time the analytization of noninteger powers was not
yet known, and the analytization was performed using an alternative way 
at the one-loop level.

In Sec.~\ref{sec:FAPT} we briefly review the FAPT model of analytization.
In Sec.~\ref{subs:pQCD} we present a two-loop pQCD calculation of the 
dressing function of
the gluon propagator in the Landau gauge.
In Sec.~\ref{subs:npQCD} we include the nonperturbative
effects, by introducing the gluon dynamical mass in the propagator, 
and by the FAPT analytization of the noninteger powers of the coupling
in the dressing function. In Sec.~\ref{sec:numres} we compare the results of
the method with the results of the unquenched ($N_f=3$) lattice data for the
propagator and with pQCD results. 
In Sec.~\ref{sec:summ} we summarize our results.

\section{Fractional Analytic Perturbation Theory (FAPT) model}
\label{sec:FAPT}

We present here a brief overview of the main ideas of the
(Fractional) Analytic Perturbation Theory ((F)APT).

\begin{figure}[htb]
\includegraphics[width=120mm]{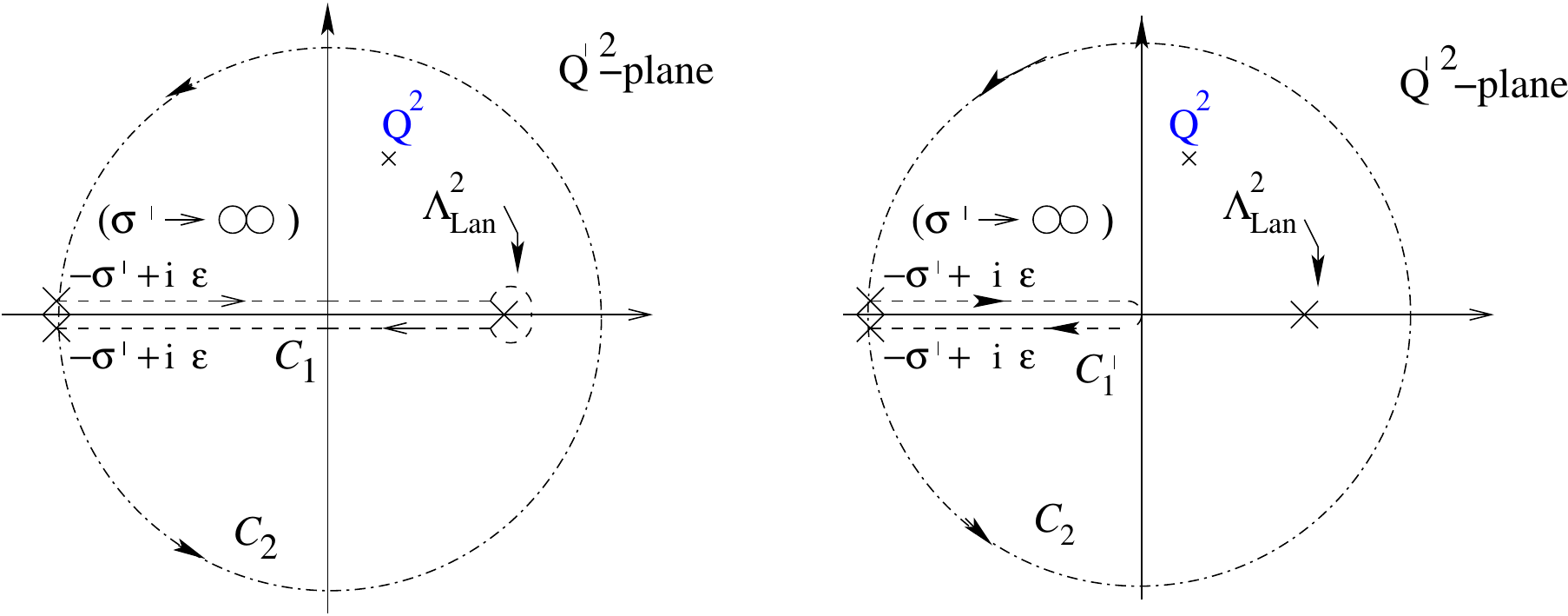}
\vspace{-0.4cm}
\caption{Left-hand Figure: the integration path for the 
integrand $a_{\rm pt}(Q'^2)/(Q'^2 - Q^2)$ leading to the 
dispersion relation (\ref{aptdisp}) for  $a_{\rm pt}(Q^2)$.  
Right-hand Figure: the integration path for the same integrand, 
leading to the dispersion relation (\ref{MAA1disp}) for the APT 
coupling $\A^{\rm (APT)}(Q^2)$.}
\label{intpath}
\end{figure}

The pQCD coupling $a(Q^2) \equiv \alpha_s(Q^2)/\pi$, in the
usual ($\MSbar$-like) schemes, is running according to the
perturbative renormalization group equation (pRGE) which
has the beta function $\beta(a)$ in the form of a truncated
power series up to $n$-loop order in $a$
\begin{eqnarray}
\frac{\partial a(Q^2; {\beta_2}, \ldots)}
{\partial \ln Q^2} 
& = &
- \sum_{j=0}^{n-1} \beta_{j} \: 
a^{j+2} (Q^2; {\beta_2}, \ldots) \ .
\label{pRGE}
\end{eqnarray}
Here, the first two beta coefficients are universal
\be 
\beta_0 = \frac{1}{4} \left( 11- \frac{2}{3} N_f \right) \ ,
\qquad
\beta_1 = \frac{1}{16} \left( 102- \frac{38}{3} N_f \right ) \ , 
\label{be0be1}
\ee
and the other coefficients $\beta_k$ ($k \geq 2$) represent
the (chosen) perturbative renormalization scheme. 
The integration of the pRGE (\ref{pRGE}) in the complex $Q^2$-plane,
with a given physical initial condition (at some high enough
positive $Q^2$), gives in general a function $a(Q^2)$ which does not
reflect the general analyticity properties of the spacelike
QCD physical quantities in the complex $Q^2$-plane: 
$a(Q^2)$ develops Landau singularities
outside the timelike region $\mathbb{C} \backslash (-\infty, -M^2_{\rm thr}]$,
most often on the positive $Q^2$ axis: $0 < Q^2 <  \Lambda^2_{\rm Lan.}$,
where $\Lambda^2_{\rm Lan.}$ ($\sim 10^{-1} \ {\rm GeV}^2$) is the
(Landau) branching point of these singularities.
When the Cauchy theorem is applied to the function $a(Q^{'2})/(Q^{'2}-Q^2)$
in the $Q^{'2}$ complex plane on an appropriate closed contour 
which avoids all the cuts but encloses the pole $Q^{'2}=Q^2$ 
(see Fig.~\ref{intpath}),  the following
dispersion relation is obtained:
\begin{equation}
a(Q^2) = \frac{1}{\pi} \int_{\sigma= - {\Lambda^2_{\rm Lan.}} - \eta}^{\infty}
\frac{d \sigma {\rho^{\rm {(pt)}}}(\sigma) }{(\sigma + Q^2)},
   \quad (\eta \to +0),
\label{aptdisp}
\end{equation}
where ${\rho^{\rm {(pt)}}}(\sigma)= {\rm Im} a(-\sigma - i \epsilon)$
is the discontinuity function of the pQCD coupling $a$
along the entire cut axis. 
The APT procedure \cite{ShS} is the elimination
of the contributions of the Landau cut 
$0 < (-\sigma) \leq \Lambda_{\rm Lan.}^2$, see Fig.~\ref{intpath}.
This gives the APT analytic analog $\A^{\rm (APT)}(Q^2)$ of $a(Q^2)$ 
\be
\A^{\rm (APT)}(Q^2) = \frac{1}{\pi} \int_{\sigma= 0}^{\infty}
\frac{d \sigma {\rho^{\rm {(pt)}}}(\sigma) }{(\sigma + Q^2)} \ .
\label{MAA1disp}
\ee
This procedure can be extended to the construction of
the analogs $\A_n^{\rm (APT)}(Q^2)$ of $n$-integer powers
$a(Q^2)^n$ \cite{MSS,Sh} and their combinations (see also \cite{KS}).
The APT analogs of general powers
$a^{\nu}$ ($\nu$ a real exponent) are obtained in the same way
\begin{equation}
{\A}^{\rm {(FAPT)}}_{\nu}(Q^2) = \frac{1}{\pi} \int_{\sigma= 0}^{\infty}
\frac{d \sigma {\rho^{\rm {(pt)}}_{\nu}}(\sigma) }{(\sigma + Q^2)} \ ,
\label{Anu}
\end{equation}
where 
\be
{\rho^{\rm {(pt)}}_{\nu}}(\sigma) = 
{\rm Im} \; a^{\nu}(Q^{'2}=-\sigma - i \epsilon) \ .
\label{rhonu}
\ee
If the underlying pQCD running coupling $a(Q^2)$ is according to the one-loop
pRGE, the corresponding explicit expressions 
for $\A_{\nu}^{\rm (FAPT)}$ exist and were obtained
and used in Ref.~\cite{BMS1}
\be
\A_{\nu}(Q^2)^{\rm (FAPT, 1-\ell.)} = \frac{1}{\beta_0^{\nu}}
\left(  \frac{1}{\ln^{\nu}(z)} -
\frac{ {\rm Li}_{-\nu+1}(1/z)}{\Gamma(\nu)} \right) \ .
\label{MAAnu1l}
\ee
Here, $z \equiv Q^2/\Lambda^2$ and 
${\rm Li}_{-\nu+1}(z)$ is the polylogarithm function of order $-\nu+1$.
Extensions of this FAPT approach to higher loops were performed 
by expanding the one-loop result in a series of derivatives
with respect to the index $\nu$ \cite{BMS1,BMS2,BMS3}. 
We refer for reviews of FAPT to Refs.~\cite{Bakulev}. 
Mathematical packages for numerical calculations in APT and FAPT
are given in Refs.~\cite{BK}. 

We will use the numerical approach (\ref{Anu}) for the calculation
of the FAPT coupling, with the underlying pQCD coupling being the
two-loop coupling \cite{Gardi:1998qr,Magradze:1998ng,Magr} 
(see also \cite{CveKon,GarKat}) 
\bea
a(Q^2) = - \frac{1}{c_1} \frac{1}{\left[
1 + W_{\mp 1}(z) \right]} \ ,
\label{aptexact}
\eea
where: $c_1 = \beta_1/\beta_0$; $Q^2=|Q^2| \exp(i \phi)$; $W_{-1}$ and $W_{+1}$
are the branches of the Lambert function
for $0 \leq \phi < + \pi$ and $- \pi < \phi < 0$, 
respectively; $z$ is defined as
\be
z =  - \frac{1}{c_1 e} 
\left( \frac{|Q^2|}{\Lambda_{\rm L.}^2} \right)^{-\beta_0/c_1} 
\exp \left( - i {\beta_0}\phi/c_1 \right) \ .
\label{zexpr}
\ee 
Here, $\Lambda_{\rm L.}$ is the Lambert QCD scale. Since we will be 
interested in the behavior of the propagator at low momenta, we will use 
for simplicity $N_f=3$ throughout, and $\Lambda_{\rm L.}(N_f=3)=0.581$ GeV.
This value corresponds to $\Lambda_{\rm L.}(N_f=5)=0.322$ GeV
(it corresponds to the $\MSbar$ scale ${\overline \Lambda}_{N_f=5} \approx 0.260$ GeV, 
used in Refs.~\cite{BMS2,BMS3,Bakulev}).

\section{Gluon propagator in the Landau gauge}
\label{sec:propth}

\subsection{Propagator in pQCD}
\label{subs:pQCD}

Gluon propagator ${\cal D}$ in the Landau gauge has the form
\be
{\cal D}^{ab}_{\mu \nu}(Q) =  \delta^{ab}
\left( g_{\mu \nu} - \frac{q_{\mu} q_{\nu}}{q^2} \right) 
D(-q^2)
\label{D}
\ee
where
\be
D(Q^2) = \frac{Z(Q^2)}{Q^2} \ ,
\label{DZ}
\ee 
and in Eq.~(\ref{DZ}) we denoted $Q^2 \equiv - q^2 \equiv - (q_0)^2 + {\vec q}^{\ 2}$ for the spacelike momenta 
($Q^2 \in \mathbb{C} \backslash (-\infty, 0]$).
Here, $Z(Q^2) \equiv Z(Q^2/\mu^2, a(\mu^2))$ is the dressing (residuum) function of 
the propagator. By the general principles of QFT, $Z$ is an analytic
(holomorphic) function of $Q^2$ in the $Q^2$-complex plane
$Q^2 \in \mathbb{C} \backslash (-\infty, 0]$, cf.~Refs.~\cite{BS,Oehme,Oehme2}.
We will assume (see the discussion in Sec.~\ref{sec:numres}) that we are in the
low-momentum region ($|Q| \equiv \sqrt{|Q^2|} \alt 1$ GeV) 
where there are three massless
active quarks ($N_f=3$).
In this Subsection we present the two-loop
pQCD expression of the dressing function $Z(Q^2)$. If the normalization 
is performed at a scale $\mu^2=Q_0^2$, we have by the renormalization
group invariance (Callan-Symanzik equation)
\be
Z(Q^{2})= \exp \left\lbrace \int^{a(Q^{2})}_{a(Q^{2}_{0})} 
\frac{\gamma(x)}{\beta(x)} dx\right\rbrace \ ,
\ee
where the anomalous dimension $\gamma$ of $Z$, and beta function $\beta$,
are taken at the two-loop level
\begin{equation}
\gamma(a)=-\left(\gamma_{0} a+\gamma_{1} a^{2}\right) \ ,
\qquad
\beta(a) =-\left(\beta_{0} a^{2}+\beta_{1} a^{3}\right) \ , 
\end{equation}
with the coefficients $\beta_0$, $\beta_1$ given in Eq.~(\ref{be0be1}),
and
\be
\gamma_0 = \frac{1}{8} \left( 13 - \frac{4}{3} N_f \right) \ ,
\qquad
\gamma_1 = \frac{1}{16} \left( \frac{531}{8} - \frac{61}{6} N_f \right) \ .
\label{g0g1}
\ee
We can rewrite
\begin{equation}
\dfrac{\gamma(a)}{\beta(a)} =
\frac{\gamma_{0}}{\beta_{0}} \dfrac{1}{a} 
\dfrac{\left[ 1+ (\gamma_{1}/\gamma_{0}) a \right]}
{ \left[ 1+(\beta_{1}/\beta_{0}) a \right] }=
\frac{\gamma_{0}}{\beta_{0}}\left[ 
\dfrac{1}{a}+\left(\dfrac{\gamma_{1}}{\gamma_{0}}-\dfrac{\beta_{1}}{\beta_{0}} \right) 
\dfrac{1}{\left( 1+ (\beta_{1}/\beta_{0}) a \right)}
\right] \ .
\end{equation}
This then gives [we denote: $a \equiv a(Q^2)$, $a_0 \equiv a(Q_0^2)$]
\bea
Z(Q^{2})&=&
\exp \left\lbrace 
\int_{a_{0}}^{a}\dfrac{\gamma(x)}{\beta(x)} dx
\right\rbrace
= \exp \left\lbrace 
\frac{\gamma_{0}}{\beta_{0}} \int_{a_{0}}^{a} 
\left[ 
\dfrac{1}{x}+\left(\dfrac{\gamma_{1}}{\gamma_{0}}-\dfrac{\beta_{1}}{\beta_{0}} \right) 
\dfrac{1}{\left( 1+ (\beta_{1}/\beta_{0}) x \right)}
\right] dx 
\right\rbrace  
\nonumber\\
&=& 
\exp\left\lbrace  
\dfrac{\gamma_{0}}{\beta_{0}} 
\ln \left( \dfrac{a}{a_{0}} \right) +
\left(\frac{\gamma_{1}}{\beta_{1}} -\frac{\gamma_{0}}{\beta_{0}}\right) 
\ln \left( \dfrac{1+ (\beta_{1}/\beta_{0}) a}{1+ (\beta_{1}/\beta_{0}) a_0} \right) 
\right\rbrace 
\nonumber\\
&=&
\left(\dfrac{a}{a_{0}}\right)^{{\gamma_{0}}/{\beta_{0}}}
\left( 
\dfrac{1+ ({\beta_{1}}/{\beta_{0}}) a}{1+ ({\beta_{1}}/{\beta_{0}}) a_{0}}
\right)^{\left(\dfrac{\gamma_{1}}{\beta_{1}}-\dfrac{\gamma_{0}}{\beta_{0}}\right)}
\label{ZQ}
\eea
This is the (formally exact) two-loop solution for the gluon
dressing function. We can expand it in powers of $a$; 
truncating at the NLO term then gives
  \begin{equation}
 Z(Q^{2})=c_{v} \left[ a^{\nu}(Q^{2})+d_{1}a^{\nu+1}(Q^{2}) \right] \ ,
\label{ZQ2l}
 \end{equation}
where $c_{v}$ is a constant (independent of $Q^2$) 
to be fixed by a renormalization condition, and we have,
according to Eq.~(\ref{ZQ})
\be
\nu = \frac{\gamma_0}{\beta_0} \ , \qquad
d_1 = \left( \frac{\gamma_1}{\beta_0} - \frac{\beta_1 \gamma_0}{\beta_0^2} \right) \ .
\label{nud1}
\ee
Numerically, $\nu=0.5$ and $d_1 \approx 0.1076$ (when $N_f=3)$.
The two-loop pQCD gluon propagator (\ref{D})-(\ref{DZ}) is then
\be
D_{\rm pt}(Q^2) = \frac{c_v}{Q^2}  \left[ a^{\nu}(Q^{2})+d_{1}a^{\nu+1}(Q^{2}) \right] \ .
\label{DQ2l}
\ee
The truncated expansion (\ref{ZQ2l}) in powers of $a(Q^2)$ was performed for two reasons.
Firstly, the analytization in the general analytic QCD frameworks
\cite{GCAK} can be performed for powers of $a(Q^2)$, 
but not for the expression of the type (\ref{ZQ}).\footnote{
Nonetheless, in (F)APT, the analytization can be performed on the
entire expressions of the type (\ref{ZQ}), i.e., $a(Q^2)^{\nu} (1 + c_1 a(Q^2))^{\mu}$,
cf.~discussion in Ref.~\cite{KS}.}
Secondly, we will be working at the two-loop level; in the
general $n$-loop case the coefficient $d_1$ of the expansion (\ref{ZQ2l}) depends 
only on the two-loop coefficients $\beta_1$ and $\gamma_1$, while the full
next coefficient $d_2$ involves also the three-loop ($\MSbar$)
coefficients $\beta_2$ and $\gamma_2$ which are not included in this work.

\subsection{Nonperturbative effects in the gluon propagator}
\label{subs:npQCD}

It is natural to consider the gluon with dynamical mass.
Various studies of the gluon propagator in the Landau gauge
using Dyson-Schwinger equations (DSEs) 
\cite{Cornwall82,CornwallPapa89,Papa90,BinosiPapa02y04,AP}
indicate that in the infrared regime (low $|Q^2|$) the 
gluon acquires, via nonperturbative effects,
a dynamical effective mass $M \approx 0.5$-$1$ GeV.
In these analyses,\footnote{ 
In Refs.~\cite{Sorella}, Gribov-Zwanziger approach extended by condensates
is applied, and qualitatively similar results for the
gluon propagator are obtained as in the mentioned DSE analyses.} 
gluon propagator in the Landau gauge tends to a nonzero constant in the
infrared, $D(Q^2) \to {\rm const.}$ when $Q^2 \to 0$ (decoupling scenarios), 
in accordance with the presently available lattice data 
\cite{lattice1,UnquenchedLatt}.\footnote{
Other DSE studies \cite{DSESme} indicate that $D(Q^2) \to 0$ (when $Q^2 \to 0$)
is also a solution to DSEs (scaling scenarios), and in Ref.~\cite{AHS} both
scenarios (scaling and decoupling) are discussed.} 
The simple version of the massive gluon propagator in the Landau gauge 
then has the form
\begin{equation}
D(Q^{2})=\frac{Z(Q^{2})}{Q^{2}+M^{2}} \ ,
\label{glM}
\end{equation}
where we assume that the dynamical mass
$M$ is $Q^2$-independent. $Z(Q^{2})$ is the dressing function
discussed in the context of pQCD in the previous Subsection.

Nonperturbative physics must affect also the powers of the
perturbative coupling $a(Q^2)$, as argued earlier.
In this work we account for such effects by
the analytization of the (two-loop) result (\ref{ZQ2l})-(\ref{nud1}) to
FAPT, by replacing $a(Q^2)^{\nu+n} \mapsto \A^{\rm (FAPT)}_{\nu+n}(Q^2)$
  \begin{equation}
 Z(Q^{2})=c_{v} \left[ \A_{\nu}^{\rm (FAPT)}(Q^{2})+
d_{1} \A_{\nu+1}^{\rm (FAPT)}(Q^{2}) \right] \ ,
\label{ZQAPT}
 \end{equation}
where $\A_{\nu}^{\rm (FAPT)}(Q^{2})$ is given by Eqs.~(\ref{Anu})-(\ref{rhonu}).

In conjunction with Eq.~(\ref{glM}), the gluon propagator function $D(Q^2)$ of Eqs.~(\ref{D})-(\ref{DZ})
is now written as
  \begin{equation}
 D(Q^{2}; c_v, M^2)=\frac{c_{v}}{Q^2 + M^2} \left[ \A_{\nu}^{\rm (FAPT)}(Q^{2})+
d_{1} \A_{\nu+1}^{\rm (FAPT)}(Q^{2}) \right] \ .
\label{DQAPT}
 \end{equation}
The introduction of the dynamical effective gluon mass in Eq.~(\ref{glM})
represents an inclusion of the nonperturbative effects which depend on
the quantity considered,\footnote{
Unlike the analytization $a^{\nu} \mapsto \A_{\nu}$ which is independent
of the quantity considered.}
which in this case is gluon propagator.
This is analogous to the approaches in 
Refs.~\cite{DeRafael,MagrDual,Milton:2001mq,mes2,Nest3,Stef}
(also: \cite{Deur,Court}) where nonperturbative
contributions are introduced in the specific considered observables.

\section{Numerical Results}
\label{sec:numres}

Before presenting our results, we note that our results should be compared
with the lattice data for the unquenched case ($N_f=3$), Ref.~\cite{UnquenchedLatt},
in the available interval of $Q \equiv \sqrt{Q^2}$: $0.1 \ {\rm GeV} < Q  \alt 10 \ {\rm GeV}$. 
This is so because the (F)APT formalism requires $N_f \geq 3$. 
Namely, the thresholds in the (F)APT formalism are understood to be implemented
in general at $Q^2=m_q^2$ ($m_q$ is the corresponding quark mass)
in the underlying pQCD coupling $a(Q^2)$, and the latter 
coupling has Landau singularities
at energies $0 < Q^2 \alt m_s^2$ ($m_s$ is the strange quark mass)
and even at $Q^2 > m_s^2$.

We work with the gluon propagator (\ref{DQAPT}), which has
FAPT-holomorphic dressing function (\ref{ZQAPT}) and a dynamical
gluon mass $M$, Eq.~(\ref{glM}). The free parameters are $c_v$ and $M$.
We are choosing certain two points of the (low-$Q$) lattice data 
and adjust the free parameters ($c_v$ and $M$) so that our curve goes 
through these two points; the two chosen points are also varied, so as
to obtain (visually) the best curve. This approach is applied in
the analytic (FAPT) and in the pQCD case.
  
In this context, we mention that a variant of APT was applied to the dressing function in
Ref.~\cite{Magradze:1998ng}, but in a more naive manner since the
FAPT approach was not known at the time; and the dynamical
gluon mass effect was not included.
\begin{figure}[htb] 
\centering\includegraphics[width=120mm]{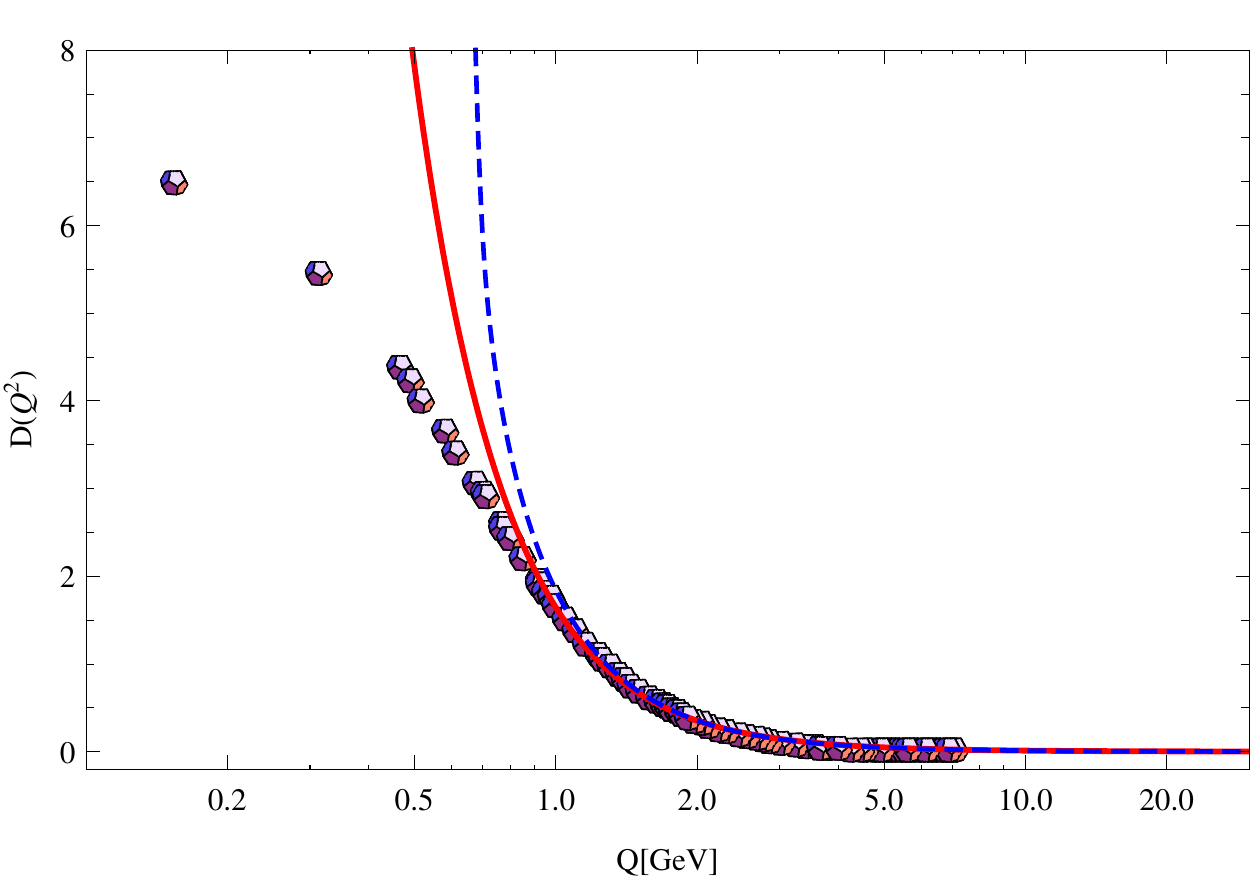}
\vspace{-0.4cm}
\caption{Analytic gluon propagator (\ref{DQAPT}),
in units of ${\rm GeV}^{-2}$,
with the dynamical effective 
gluon mass parameter $M=0$ and $c_v=5.10$ (continuous line),
in comparison with unquenched lattice data taken from
Ref.~\cite{UnquenchedLatt}, where $N_f=2+1$. Further, the (two-loop) pQCD
result (\ref{DQ2l}) is presented as well, with $c_v=4.55$ (dashed line).}
\label{GPAPTfig}
 \end{figure}

\begin{figure}[htb] 
\centering\includegraphics[width=120mm]{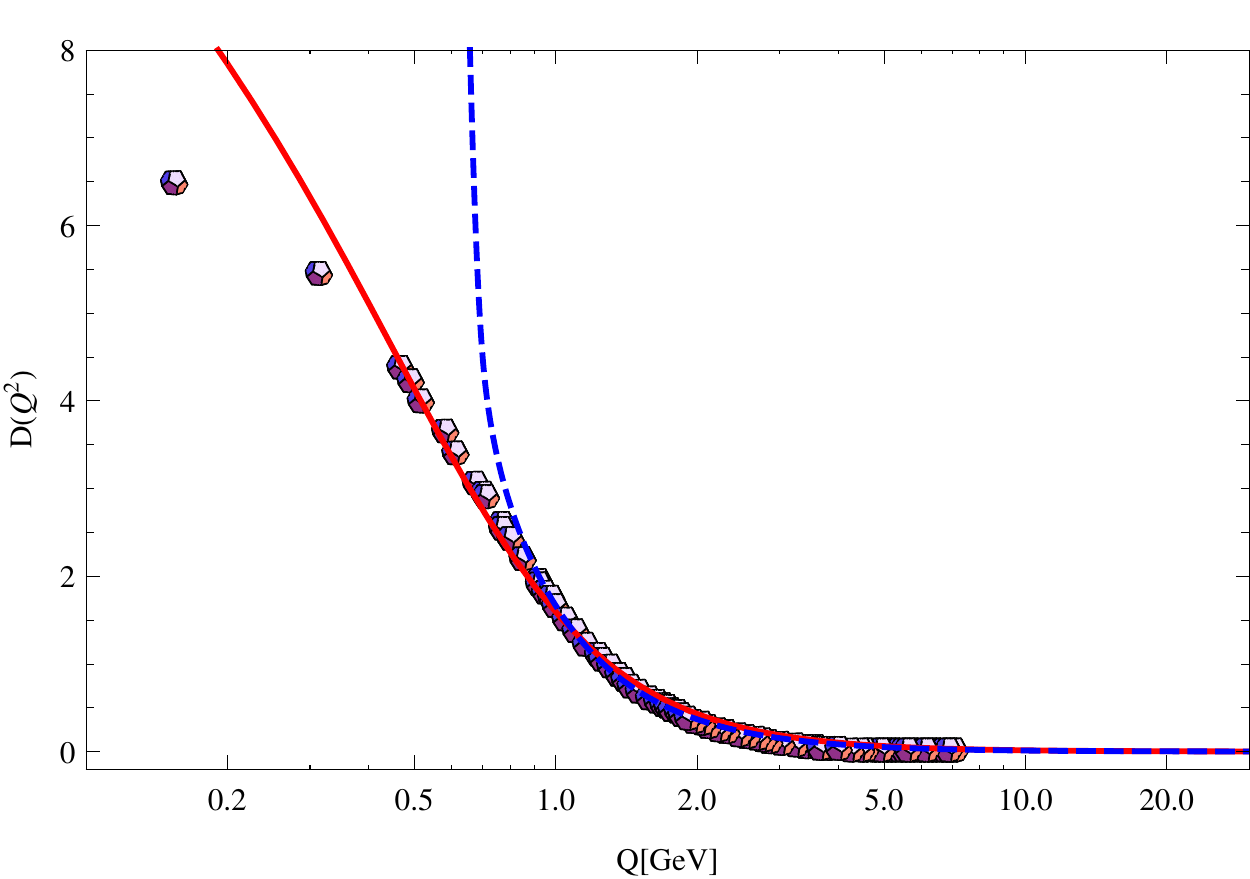}
\vspace{-0.4cm}
\caption{The same as in Fig.~\ref{GPAPTfig}, but now
the dynamical effective gluon mass parameter $M$ is nonzero: 
the analytic propagator
(\ref{DQAPT}) (continuous line) has $M^2=0.382 \ {\rm GeV}^2$ and
$c_v=6.81$. The pQCD result (dashed line), i.e., 
Eq.~(\ref{glM}) with (\ref{ZQ2l}), has $M^2 =0.211 \ {\rm GeV}^2$ 
and $c_v = 4.98$.}
\label{GPMAPTfig}
 \end{figure}

In Fig.~\ref{GPAPTfig}, we compare with the lattice data at low
positive $Q^2$ ($Q \equiv \sqrt{Q^2}$) when fixing $M=0$
in $D(Q^2)$  of Eq.~(\ref{DQAPT}), while
the other parameter $c_v$ is fixed by a lattice point.  We include also
a (two-loop) pQCD result (\ref{DQ2l}), where in $a(Q^2)$ we use the same
Lambert scale: $\Lambda_{\rm L.}(N_f=3)=0.581$ GeV.

In Fig.~\ref{GPMAPTfig} both parameters, $M$ and $c_v$, are varied in
Eq.~(\ref{DQAPT}) so as to get the best possible
agreement with the lattice data. For additional comparisons, a representative
pQCD curve with $M \not= 0$ [i.e., Eq.~(\ref{glM}) with (\ref{ZQ2l})] is
included, where we use the aforementioned Lambert scale.

In Fig.~\ref{GPAPTfig} we see that the analytization alone (and $M=0$) gives us
results which are good down to approximately $0.8$ GeV; this is 
an improvement with respect to pQCD. 
But we wanted to go beyond and see what is the effect of including a (small) 
dynamical effective mass $M$ of gluon, Eqs.~(\ref{glM}) and (\ref{DQAPT}).
We find that the best mass parameter is $M^2 \approx 0.382 \ {\rm GeV}^2$, 
see Fig.~\ref{GPMAPTfig}.
It turns out that this value is consistent with 
the values of $M$ obtained in Refs.~\cite{OB}.\footnote{In Refs.~\cite{OB},
$M(Q^2)$ and $Z(Q^2)$ were considered as $Q^2$-dependent functions
with specific Ans\"atze, and the resulting propagator was fitted
to the lattice results.}

We see from Fig.~\ref{GPMAPTfig} that the massive FAPT-analytic version is 
applicable for momenta down to $Q \approx 0.4$ GeV 
($Q^2 \approx 0.15 \ {\rm GeV}^2$),
but not below that. This is consistent with the conclusions about the
applicability of the APT approach in Bjorken Polarized Sum Rule at low
$Q$, Ref.~\cite{Khandramai:2011zd}, where the authors included
other nonperturbative effects via higher-twist terms.

We note that in our approach, when we want to reproduce lattice results 
for lower momenta $Q$, the mass parameter $M$ is getting bigger and 
is accompanied with a worse behavior at higher $Q$ values. 
Therefore, we intend to improve this approach in the future, by
using a $Q^2$-dependent dynamical mass $M(Q^2)$ of the gluon.

If we performed our analysis with the one-loop FAPT, this would affect 
in the present approach only the dressing function $Z(Q^2)$ and not the mass
$M^2$, i.e., in Eq.~(\ref{ZQAPT}) we would have only one term.
This would give us almost the same result as in the two-loop case of FAPT, and
the resulting curves would be almost indistinguishable from the solid curves in 
Figs.~\ref{GPAPTfig},\ref{GPMAPTfig}. This is due to the small value of $d_1 \approx 0.1076$
and due to the very good hierarchy of FAPT couplings: $|\A_{\nu+1}^{\rm (FAPT)}(Q^{2})|
\ll |\A_{\nu}^{\rm (FAPT)}(Q^{2})|$, true even at low $|Q^2|$, cf.~Refs.~\cite{Bakulev}.
In the context of DSEs, the effect of two-loop effects was investigated in 
Ref.~\cite{Huber}.

\section{Summary}
\label{sec:summ}

In this work, we evaluated the gluon propagator in the Landau gauge at low spacelike
momenta $Q^2$.
We used  the two-loop solution of the Callan-Symanzik equation
for the dressing function $Z(Q^2)$. The nonperturbative effects were incorporated in
the form of the analytization procedure $a^{\nu}(Q^2) \mapsto \A_{\nu}(Q^2)$ for the
(noninteger) powers of the QCD coupling $a(Q^2)$, within 
Fractional Analytic Perturbation Theory (FAPT) with $N_f=3$; 
and by incorporation of a constant dynamical effective gluon mass $M$ 
in the propagator. The obtained expression Eq.~(\ref{DQAPT}) has
two free parameters: the normalization constant $c_v$ and the dynamical gluon mass $M$.
We compared the obtained results with the unquenched lattice results ($N_f=3$)
at low positive $Q^2$. 
In comparison with pQCD results, the analytization clearly improved the behavior 
of the propagator at low $Q$. The additional introduction of the
dynamical effective gluon mass further improved the low-$Q$ behavior.
We used the FAPT with $N_f=3$, because in FAPT it is apparently not possible to define
unambiguously the theory for $N_f < 3$.

The main results can be summarized as follows:
\begin{enumerate}
\item We performed an analytization procedure (numerical FAPT) 
for the dressing function of gluon propagator in the Landau gauge. The
dressing function was obtained from the two-loop 
Callan-Symanzik equation, and this (pQCD) procedure introduces one free parameter
$c_v$ which is an overall normalization constant.
\item In addition, we introduced a (constant) dynamical effective gluon mass $M$ 
in the propagator, as suggested by various DSE studies of gluon propagator
in the Landau gauge.
 \item We compared the obtained results 
with the unquenched lattice data of the propagator,
by varying the free parameters $c_v$ and $M$.
\item
In the nonmassive ($M=0$) case, the analytic gluon propagator 
is in agreement with the unquenched lattice data down to 
$Q^2 \approx  0.6 \ {\rm GeV}^2$ ($Q \equiv \sqrt{Q^2} \approx 0.8$ GeV),
while the massive ($M >0$) analytic gluon propagator agrees
with the lattice data down to $Q^2 \approx  0.15 \ {\rm GeV}^2$ ($Q \approx 0.4$ GeV).
 \item The values that we found for our fit are 
$c_v=5.10$ for the nonmassive case; and $c_v=6.81$ and $M^2=0.382 \ {\rm GeV}^2$
($M \approx 0.62$ GeV) for the massive case, 
where this value of the dynamical effective gluon mass $M$ is similar to the 
values found in the literature.
\end{enumerate}

We intend to continue this work in various directions: perform the
analytization procedure within other QCD models such as the two-delta analytic
QCD model of Ref.~\cite{2danQCD} and the analytic QCD models with
effective mass in the coupling 
(cf.~Refs.~\cite{Simonov,BadKuz,BKS,KKSh,Shirkovmass,Luna}); 
allow $Q^2$-dependence in the dynamical effective
gluon mass $M$; and compare with DSE results (numerically and theoretically).

\begin{acknowledgments}
\noindent
This work was supported in part by FONDECYT (Chile) Grant No.~1130599 (G.C.),
and in part by CONICYT (Chile) Beca Mag\'{\i}ster Nacional No.~22121757 (P.A.).
\end{acknowledgments}

  \end{document}